\documentclass[twocolumn,showpacs,preprintnumbers,amsmath,amssymb,APSl,prd,nofootinbib,superscriptaddress]{revtex4-1}

\usepackage{dcolumn}
\usepackage{bm}
\usepackage{ifpdf}
\usepackage{hyperref}
\usepackage{bm}
\usepackage{xcolor,color,graphicx,graphics}
\usepackage[spanish,english]{babel}
\usepackage[latin1]{inputenc}
\usepackage[OT1]{fontenc}
\usepackage{latexsym,amssymb,amsmath,amsfonts}
\usepackage{makeidx}
\usepackage{epsfig,subfigure}
\usepackage{natbib}
\usepackage{epstopdf}
\usepackage{mathrsfs}
\usepackage{hyperref}
\hypersetup{colorlinks=true, linkcolor=blue, citecolor=green}
\usepackage{enumerate}
\usepackage{tikz}

\definecolor{red}{rgb}{1,0,0}

\def\+{^\dagger}

\def\<{\leftarrow}
\def\>{\rightarrow}

\def\({\left(}
\def\){\right)}

\newcommand{\bi}{\begin{itemize}} 				\newcommand{\ei}{\end{itemize}}
\newcommand{\benu}{\begin{enumerate}} 		\newcommand{\enu}{\end{enumerate}}
\newcommand{\bd}{\begin{dinglist}{0}}     \newcommand{\ed}{\end{dinglist}}
\newcommand{\bfig}{\begin{figure}[htbp]}  \newcommand{\efig}{\end{figure}}
        			
\newcommand{\bc}{\begin{center}} 				  \newcommand{\ec}{\end{center}}
\newcommand{\be}{\begin{equation}} 				\newcommand{\ee}{\end{equation}}
\newcommand{\bsub}{\begin{subequations}}  \newcommand{\esub}{\end{subequations}}
\newcommand{\ben}{\begin{eqnarray}} 			\newcommand{\een}{\end{eqnarray}}
\newcommand{\ba}[1]{\begin{array}{#1}} 		\newcommand{\ea}{\end{array}}
\newcommand{\bea}{\begin{equation}\begin{array}{rcl}}
\newcommand{\eea}{\end{array}\end{equation}}

\definecolor{purple}{rgb}{1,0,1}
\definecolor{lime}{HTML}{A6CE39} 

\newcommand{\orcidicon}{%
	\begin{tikzpicture}
	\draw[lime, fill=lime] (0,0)
		circle [radius=0.16]
		node[white] {{\fontfamily{qag}\selectfont \tiny ID}};
	\draw[white, fill=white] (-0.0625,0.095)
		circle [radius=0.007];
	\end{tikzpicture}	\hspace{-2mm}
}
\newcommand\orcidCabral{{\href{https://orcid.org/0000-0003-2124-5894}{\orcidicon}}}
\newcommand\orcidLobo{{\href{https://orcid.org/0000-0002-9388-8373}{\orcidicon}}}
\newcommand\orcidDiego{{\href{https://orcid.org/0000-0003-3984-9864}{\orcidicon}}}

\begin{document}

\title{The cosmological principle in theories with torsion: the case of Einstein-Cartan-Dirac-Maxwell gravity}

\author{Francisco Cabral\orcidCabral\!\!}
\email{ftcabral@fc.ul.pt}
\affiliation{Instituto de Astrof\'{\i}sica e Ci\^{e}ncias do Espa\c{c}o, Faculdade de
Ci\^encias da Universidade de Lisboa, Edif\'{\i}cio C8, Campo Grande,
P-1749-016 Lisbon, Portugal}
\author{Francisco S. N. Lobo\orcidLobo\!\!} \email{fslobo@fc.ul.pt}
\affiliation{Instituto de Astrof\'{\i}sica e Ci\^{e}ncias do Espa\c{c}o, Faculdade de
Ci\^encias da Universidade de Lisboa, Edif\'{\i}cio C8, Campo Grande,
P-1749-016 Lisbon, Portugal}
\author{Diego Rubiera-Garcia\orcidDiego\!\!} \email{drubiera@ucm.es}
\affiliation{Departamento de F\'isica Te\'orica and IPARCOS, Universidad Complutense de Madrid, E-28040
Madrid, Spain}

\date{\today}

\begin{abstract}
We address the implementation of the cosmological principle, that is, the assumption of  homogeneity and isotropy in the spatial distribution of matter in the Universe, within the context of Einstein-Cartan theory including minimal couplings of both Dirac  and Maxwell fields to torsion. This theory gives rise to new physical effects in environments of high spin densities while leaving the vacuum dynamics unaffected. Our approach is to impose the cosmological principle from the onset to the geometrical degrees of freedom (metric and torsion functions), which constrains the torsion components and the corresponding correction terms in the Friedmann-like equations and in the resulting fermionic and bosonic (non-linear) dynamics. We derive the corresponding cosmological dynamics for the geometrical and  matter degrees of freedom and discuss the validity of this approach.

\end{abstract}

\maketitle

\section{Introduction}

In the last decades, cosmology has undergone a metamorphosis driven by the development of new technologies and the establishment of ambitious international collaborations, such as  Planck \cite{Aghanim:2018eyx}, Euclid \cite{Amendola:2012ys}, the LIGO/Virgo collaboration \cite{Abbott:2016blz}, or LISA \cite{Barausse:2020rsu} to name but a few. The access to a vast amount of information has allowed to shape the cosmological concordance $\Lambda$CDM model, anchored in Einstein's General Theory of Relativity (GR) \cite{Will:2014kxa}, and including an early inflationary phase, a cold dark matter contribution, and a tiny cosmological constant, which has successfully met all observational tests \cite{Bull:2015stt}. This model, however, has also faced some fundamental difficulties, such as: the nature of the dark matter components, which despite intense observational searches remains unknown \cite{DMS1,DMS2,DMS3}; the degeneracy in the inflationary models  and the search for the inflaton \cite{Lidsey:1995np}; and the nature of the exotic fluid driving the late-time accelerated expansion of the universe \cite{Copeland:2006wr,Frusciante:2019xia}. In view of these difficulties, a popular trend in the community is to modify the theoretical description of the gravitational dynamics. Consequently, a large number of such modified gravitational models have been proposed based on different principles, see \cite{DeFelice:2010aj,Olmoreview,CLreview,NOOreview,BeltranJimenez:2017doy,Heisenberg:2018vsk,Cambridgebook} for several reviews on the topic.

One of the main difficulties faced by these proposals is to ensure their consistency with the predictions of the $\Lambda$CDM model (and, more generally, of GR itself) in those scales where it is known to be observationally successful, which leaves little room for drastic alterations of the main principles upon which GR is built. A path to consistently explore the cosmological dynamics beyond that described in the $\Lambda$CDM model is the consideration of theories with torsion, a topic which has raised a renewed interest in the last few years \cite{Shie:2008ms,Harko:2014sja,Harko:2014aja,Nikiforova:2016ngy,Valdivia:2017sat,Banerjee:2018yyi,Barrow:2019bvx,Pereira:2019yhu,delaCruz-Dombriz:2018aal,Cembranos:2018ipn}. The canonical model at this regard is the Einstein-Cartan-Sciama-Kibble (ECSK) theory, where the spacetime is endowed with a Riemann-Cartan (RC) geometry with curvature and torsion (for a quick glance on the main elements of this theory, see for instance \cite{Trautman:2006fp}). The torsion tensor (the antisymmetric part of the affine connection) does not propagate in vacuum, which ensures its consistency with weak-field limit observations. However, the theory induces new physical effects in regions where high spin densities are present, such as in the early Universe, due to modified Einstein-like equations and extended dynamics governing the matter fields, as has been investigated in a number of works \cite{Poplawski:2011jz,Unger:2018oqo,Kranas:2018jdc,Poplawski:2010kb,Ivanov:2016xjm,Razina:2010bj,Palle:2014goa, Poplawski:2012qy,Xue:2008qs,Vakili:2013fra}.

In a previous work, we considered ECSK theory with minimal couplings of  fermions (Dirac) and bosons (Maxwell) to the RC geometry \cite{Cabral:2019gzh}. The resulting Einstein-Cartan-Dirac-Maxwell (ECDM) theory leads to non-linear generalized Dirac-Hehl-Data and non-linear electromagnetic equations, with non-minimal interactions between fermionic and bosonic fields. As it is well known, the minimal coupling of Maxwell fields to torsion breaks the electromagnetic $U(1)$ gauge symmetry, which can therefore be linked to a cosmological phase transition during the torsion era. The cosmological dynamics of this model harbours some surprises, including the existence of cosmological bounces replacing the Big Bang singularity, or the existence of cyclic Universes linking an expansion phase with an accelerated contraction one, as well as including an early acceleration phase \cite{Cabral:2020mst}.

The main aim of the present work is to determine the restrictions to the dynamics of the ECDM model from the cosmological principle, namely, the assumption on the homogeneity and isotropy in the spatial distribution of matter at large scales. In fact, we refer the interested reader to Ref. \cite{Tsamparlis:1981xm}, for one of the first works to consider the form of torsion in spatially homogeneous and isotropic spacetimes.
Indeed, when dealing with cosmological settings in models with torsion, typically one makes use of the Weyssenhof spin fluid which, although it is known not to be fully compatible with the cosmological principle, it is possible to invoke macroscopic averaging arguments to avoid inconsistencies (see for instance the discussion of \cite{Blagojevic:2013xpa}). Similar macroscopic procedures have to be taken into account when one deals with fundamental fermionic (Dirac) degrees of freedom, in order to obtain a (perfect) fluid description. The Weyssenhof spin fluid can thus be seen as a valid approximation for a classical description of a fluid with macroscopic spin due to fermionic matter.

Here we consider an alternative approach on implementing the cosmological principle in models with torsion. We consider the appropriate fundamental fields representing fermionic matter (Dirac spinors) and bosonic (four-vector) fields by assuming that they couple minimally to torsion at the microscopic level. We then drop any assumptions about the source of torsion, which can be seen as an external function enriching the background geometry, and impose the cosmological principle from the onset. This will introduce certain conditions on the torsion tensor components and on the corresponding (correction) terms in the dynamical equations governing the geometry and the matter fields. Determining the correct shape of such corrections under a rigorous implementation of the cosmological principle is expected to have relevant consequences in the physics of the early and late-time Universe within this kind of models. We finally consider the Cartan equations in this context and explore the resulting dynamics, upon replacing the allowed torsion function (compatible with the cosmological principle) as a function of the fundamental matter fields.

This paper is organized as follows: in Sec. \ref{secII} we introduce Einstein-Cartan theory and consider the minimal coupling of Dirac and Maxwell fields to the RC geometry. In Sec. \ref{secIII} we implement the cosmological principle in Einstein-Cartan-Dirac-Maxwell gravity in a FLRW background. Finally, in Sec. \ref{secIV} we discuss the validity of our approach and mention some possible limitations to the application of the cosmological principle in theories with torsion. As usual, and unless stated otherwise, greek indices run on $\mu =0,1,2,3$ and latin ones on $a= 1,2,3$.

\section{Einstein-Cartan-Dirac-Maxwell theory}\label{secII}

\subsection{The ECSK action and Cartan equations}

The Einstein-Cartan theory is given by the following action
\begin{equation} \label{eq:actionEC}
S_{\rm EC}=\dfrac{1}{2\kappa^2} \int d^4x \sqrt{-g}R(\Gamma) + \int d^4x \sqrt{-g}\,\mathcal{L}_{m}\,,
\end{equation}
where $\kappa^2=8\pi G$ and $g$ is the determinant of the spacetime metric $g_{\mu\nu}$. This action is defined in a RC spacetime with curvature and torsion, where the curvature scalar $R=g_{\mu\nu}R^{\mu\nu}$ is constructed out of the Ricci tensor $R_{\mu\nu}(\Gamma) \equiv {R^\alpha}_{\mu\alpha\nu}(\Gamma)$. Torsion is given by $T^{\alpha}{}_{\mu\nu} \equiv \Gamma^{\alpha}{}_{[\mu\nu]}$, where the affine (Cartan) connection $\Gamma_{\mu\nu}^{\lambda}$ can be split in two pieces as
\begin{equation}
\Gamma_{\mu\nu}^{\lambda}=\tilde{\Gamma}_{\mu\nu}^{\lambda}+K_{\mu\nu}^{\lambda} \ ,
\end{equation}
where the term $\tilde{\Gamma}_{\mu\nu}^{\lambda}$ is the Levi-Civita part associated to (Riemaniann) curvature, and $K_{\alpha\mu\nu}$ is the contortion tensor related to torsion via the formula
\begin{equation} \label{eq:contdef}
K_{\alpha\mu\nu}=T_{\alpha\mu\nu}+2T_{(\mu\nu)\alpha} \,.
\end{equation}
Note that the contortion tensor is antisymmetric in the first two indices, $K_{\alpha\beta\gamma}=-K_{\beta\alpha\gamma}$, and, by construction, it satisfies $K_{\alpha[\beta\gamma]}=T_{\alpha\beta\gamma}$. As for the matter Lagrangian $\mathcal{L}_m=\mathcal{L}_{m}(g_{\mu\nu},\Gamma,\psi_m)$, it depends on the metric, the matter fields $\psi_m$, and on the contortion via the covariant derivatives in the kinetic terms.

We point out that the action (\ref{eq:actionEC}), though formally identical to the Einstein-Hilbert action of GR, possesses the richer spacetime RC geometry which introduces torsion-induced corrections to both the gravitational Einstein-like equations and to the dynamical equations for the matter fields. The derivation of both sets of equations proceeds by independent variation of the above action with respect to the tetrads $\vartheta^{a}_{\;\mu}$ (or metric), the Lorentz spin connection $w^{a}_{\;b\mu}$ (or contortion) and the matter fields. The variables $(\vartheta^{a}_{\;\mu},w^{a}_{\;b\mu})$ are the appropriate gauge potentials representing the gravitational degrees of freedom within a Poincar\'{e} gauge description of gravity (as is the ECSK) and are in line with the Einstein-Palatini metric-affine approach\footnote{In the set of variables $(\vartheta^{a}_{\;\mu},w^{a}_{\;b\mu})$, the latin indices are symmetry/group indices while the greek indices are spacetime indices.}. In what follows all quantities with a tilde represent the corresponding expressions computed in GR (curved, pseudo-Riemann spacetime), which will facilitate the comparison of our results with those of GR.

A key element in ECDM theory is that torsion does not propagate and obeys the Cartan equations:
\begin{equation} \label{eq:Careq}
T^{\alpha}_{\;\beta\gamma}+\delta^{\alpha}_{\beta}T_{\gamma}-\delta^{\alpha}_{\gamma}T_{\beta}=\kappa^2 s^{\alpha}_{\;\beta\gamma} \ ,
\end{equation}
where $s^{\alpha\mu\nu}\equiv \delta \mathcal{L}_{m}/\delta K_{\mu\nu\alpha}$ is the spin tensor of matter, with dimensions of energy/area or spin/volume. Cartan's equations (\ref{eq:Careq}) allow us to replace the torsion terms as functions of the matter fields.
Although we will proceed next without taking into account the Cartan equations (and the related non-linearities in the matter sector induced by torsion), we will consider these equations in the final part of this paper.

\subsection{Matter fields in RC spacetime}

Let us begin with the main elements regarding the matter Lagrangians in the RC geometry, and we refer the reader to Ref. \cite{Cabral:2019gzh} for further details. We consider classical fermionic (four-spinor) and bosonic (four-vector) fields minimally coupled to the RC spacetime geometry, given by the  Lagrangian density
\begin{equation}
\mathcal{L}_{m}=\mathcal{L}_{\rm D}+\mathcal{L}_{\rm M}+j^{\mu}A_{\mu}.
\label{MattLagrang}
\end{equation}
The first term is the Dirac Lagrangian in a RC spacetime, given by
\begin{equation}
\mathcal{L}_{\rm D}=\dfrac{i\hbar}{2}\left(\bar{\psi}\gamma^{\mu}D_{\mu}\psi-(D_{\mu}\bar{\psi})\gamma^{\mu}\psi\right)-m\bar{\psi}\psi \ ,
\label{DiracLag0}
\end{equation}
for spinors $\psi$ and their complex conjugates $\bar{\psi}$, while the covariant derivatives are defined as
\begin{eqnarray}
D_{\mu}\psi&=&\tilde{D}_{\mu}\psi+\dfrac{1}{4}K_{\alpha\beta\mu}\gamma^{\alpha}\gamma^{\beta}\psi  \\
D_{\mu}\bar{\psi}&=&\tilde{D}_{\mu}\bar{\psi}-\dfrac{1}{4}K_{\alpha\beta\mu}\bar{\psi}\gamma^{\alpha}\gamma^{\beta} \ ,
\end{eqnarray}
where we have introduced $D_{\mu}$ and $\tilde{D}_{\mu}$ as the (Fock-Ivanenko) covariant derivatives built with the Cartan connection and the Levi-Civita connection, respectively, and $\gamma^{\mu}$ are the induced Dirac-Pauli matrices which obey $\left\lbrace \gamma^{\mu},\gamma^{\nu}\right\rbrace =2g^{\mu\nu}I$, where $I$ is the $4\times4$ unit matrix. We can furthermore rewrite Eq. (\ref{DiracLag0}) as
\begin{equation}
\label{DiracLagv2}
\mathcal{L}_{\rm D}=\tilde{\mathcal{L}}_{\rm D}+
K_{\alpha\beta\mu}s^{\mu\alpha\beta}_{D} \ ,
\end{equation}
where $\tilde{\mathcal{L}}_{\rm D}$ is the Dirac Lagrangian in a curved Riemannian spacetime, and $s_{\;\;\alpha\beta\gamma}^{D}=\delta \mathcal{L}_{m}/\delta K^{\beta\gamma\alpha}=\dfrac{1}{2}\epsilon_{\alpha\beta\gamma\lambda}\breve{s}^{\lambda}$ is the totally antisymmetric Dirac's spin tensor, expressed in terms of the axial (spin) vector $\breve{s}^{\lambda}=\dfrac{\hbar}{2}\bar{\psi}\gamma^{\lambda}\gamma^{5}\psi$. Another possible form of the Dirac Lagrangian (\ref{DiracLagv2}) is given by
\begin{equation}
\label{DiracLag3}
\mathcal{L}_{\rm D}=\tilde{\mathcal{L}}_{\rm D}+3\breve{T}^{\lambda}\breve{s}_{\lambda} \ ,
\end{equation}
where we have introduced the axial vector part of torsion:
\begin{equation} \label{eq:axialpart}
\breve{T}^{\lambda}\equiv \dfrac{1}{6}\epsilon^{\lambda\alpha\beta\gamma}T_{\alpha\beta\gamma} \ .
\end{equation}
Expression (\ref{DiracLag3}) highlights the fact that in the minimal coupling to the RC geometry Dirac spinors only couple to the axial torsion vector.

The second term, $\mathcal{L}_{\rm M}$, in the matter action (\ref{MattLagrang}) represents massless (four-vector) bosons corresponding to the generalization of the Maxwell Lagrangian to RC spacetime:
\begin{equation}\label{eq:Maxfull}
\mathcal{L}_{\rm M}=\frac{\lambda}{4}F_{\mu\nu}F^{\mu\nu} \ ,
\end{equation}
where $\lambda$ is a coupling parameter and the generalized Faraday tensor is defined as
\begin{equation}
\label{newFaraday}
F_{\mu\nu}\equiv \nabla_{\mu}A_{\nu}-\nabla_{\nu}A_{\mu}=\tilde{F}_{\mu\nu}+2K^{\lambda}_{\;\;[\mu\nu]}A_{\lambda} \,,
\end{equation}
and $\nabla_\mu$ is the covariant derivative in a RC spacetime constructed with the Cartan connection, while $\tilde{F}_{\mu\nu}=\partial_{\mu}A_{\nu}-\partial_{\nu}A_{\mu}$ is the usual Faraday tensor when torsion is neglected. We can rewrite this expression using the contortion tensor as
\begin{equation}
\label{BosonLagcorr}
\mathcal{L}_{\rm M}=\tilde{\mathcal{L}}_{\rm M}+\lambda \left(K^{\lambda[\mu\nu]}K^{\gamma}{}_{[\mu\nu]}A_{\gamma}+K^{\lambda[\mu\nu]}\tilde{F}_{\mu\nu} \right)A_{\lambda} \ ,
\end{equation}
and the second term above (which is a correction with regards to the usual Maxwell Lagrangian), in terms of torsion is given by
\begin{equation}
\mathcal{L}^{\rm M}_{\rm corr}\equiv \lambda \left(T^{\lambda\mu\nu}T^{\gamma}_{\;\;\mu\nu}A_{\gamma}+T^{\lambda\mu\nu}\tilde{F}_{\mu\nu} \right)A_{\lambda} \ .
\end{equation}
Note that if one regards the photon-Maxwell field as the massless spin-1 within the $U(1)$ gauge theory, then strictly speaking, the spin-1 bosonic field we are describing in this paper, as well as in \cite{Cabral:2019gzh,Cabral:2020mst}, should not be regarded as the Maxwell field interacting with torsion, but rather as some other fundamental field which behaves in a similar way as an extended Proca-field (with the background torsion providing an effective mass). If torsion is ``switched off'', then it explicitly behaves as the photon-Maxwell field. In this sense, the cosmology of the early Universe is the appropriate scenario to induce such a type of phase transitions.

Finally, the last term, $j^{\mu}A_{\mu}$, in the matter Lagrangian (\ref{MattLagrang}), where $j^{\mu}=q\bar{\psi}\gamma^{\mu}\psi$ is the Dirac charge current four-vector, represents the usual minimal coupling between the charged fermionic and massless bosonic fields.

\section{Cosmological principle and the $U(1)$ symmetry breaking} \label{secIII}

To implement the cosmological principle from the onset, one needs to consider the six Killing vectors $\xi$ related to the isometries of the maximally symmetric spatial hypersurfaces, which imply the following Lie derivatives along the directions of such vectors (see \cite{Blagojevic:2013xpa} and references therein)
\begin{equation}
L_{\xi}g_{\mu\nu}=0, \qquad L_{\xi}T^{\mu}_{\alpha\beta}=0 \ .
\end{equation}
Therefore the metric is of Friedmann-Lemaitre-Robertson-Walker (FLRW) type, namely,
\begin{equation}
ds^2=dt^2 -a^2(t)\left(\frac{dr^2}{1-k r^2}+r^2 d\Omega^2 \right) \ ,
\end{equation}
where $a(t)$ is the scale factor, $k=-1,0,1$ denotes the curvature of space, and the only non-vanishing components of the torsion tensor obey
\begin{equation} \label{eq:ansatz}
T_{abc}=f(t)\epsilon_{abc}, \qquad T^{a}_{\;\;b0}= h(t) \delta^{a}_{b} \ ,
\end{equation}
where $f(t)$ and $h(t)$ are arbitrary functions of time, while $\epsilon_{abc}$ and $\delta^{a}_{b}$ are the three-dimensional Levi-Civita and Kronecker symbols, respectively.
As a consequence, from the definition of the contortion tensor (\ref{eq:contdef}), we obtain
\begin{equation} \label{eq:Kansatz}
K_{abc}=f(t)\epsilon_{abc}\,, \quad K_{0ab}=-K_{a0b}= 2h(t)g_{ab} \,,
\end{equation}
while any other component vanishes. This is valid for any gravity theory with a RC spacetime if one imposes the cosmological principle to the torsion tensor. Let us now consider the bosonic and fermionic sectors in this context.

\subsection{Bosonic (vector) fields}

Writing the electromagnetic piece of the Lagrangian density in (\ref{MattLagrang}) as $\mathcal{L}^{\rm EM}=\mathcal{L}^{\rm M}+j^{\lambda}A_{\lambda}$, with $\mathcal{L}^{\rm M}$ given by Eq. (\ref{BosonLagcorr}), the correction term (implementing the $U(1)$-symmetry breaking) with the  ansatz (\ref{eq:ansatz}) becomes
\begin{eqnarray} \label{eq:Lagcom}
\mathcal{L}^{\rm M}_{\rm corr}&=&\lambda\Big[2\left(f^{2}(t)+h^{2}(t)\right)A^{a}A_{a}\nonumber \\
&&+f(t)\epsilon^{abc}\tilde{F}_{bc}A_{a}-2h(t)\tilde{F}_{0b}A^{b}\Big] \,,
\end{eqnarray}
with the four-potential $A_{\mu}=(\phi,\vec{A})$.
On the other hand, by varying the full Lagrangian in (\ref{MattLagrang}) with respect to $A_{\mu}$ one gets
\begin{equation}
\nabla_{\mu}F^{\mu\nu}=\lambda^{-1}j^{\nu} \,,
\end{equation}
which can be conveniently rewritten as
\begin{equation} \label{eq:ECDEem}
\qquad \tilde{\nabla}_{\mu}\tilde{F}^{\mu\nu}=\lambda^{-1}(j^{\nu}+J^{\nu}) \ ,
\end{equation}
where we have defined the torsion-induced four-current
\begin{eqnarray}
\label{newcurrentgeneral}
J^{\nu}&=&-\lambda\Big[2(K^{\nu}_{\;\;\lambda\mu}K^{\gamma[\mu\lambda]}+K_{\lambda}K^{\gamma[\lambda\nu]})A_{\gamma}\,
	 \\
&&+K^{\nu}_{\;\;\lambda\mu}\tilde{F}^{\mu\lambda}+K_{\lambda}\tilde{F}^{\lambda\nu}
+2\tilde{\nabla}_{\mu}\left(K^{\gamma[\mu\nu]}A_{\gamma}\right)\Big] \ , \nonumber
\end{eqnarray}
with $K_{\lambda}\equiv K^{\alpha}_{\;\;\lambda\alpha}$. 

The generalized current can also be written as
\begin{eqnarray} \label{eq:Jcomp}
J^{\nu}&=&-\lambda\Big[2(T^{\nu}_{\;\;\lambda\mu}T^{\gamma\mu\lambda}+2T_{\lambda}T^{\gamma\lambda\nu})A_{\gamma}\,
\nonumber \\
&&+T^{\nu}_{\;\;\lambda\mu}\tilde{F}^{\mu\lambda}+2T_{\lambda}\tilde{F}^{\lambda\nu}
+2\tilde{\nabla}_{\mu}(T^{\gamma\mu\nu}A_{\gamma})\Big] \ ,
\end{eqnarray}
where we have used the fact that contortion is antisymmetric in the first two indices and also that $K^{\nu}_{\;\;[\lambda\mu]}=T^{\nu}_{\;\;\lambda\mu}$ and $K_{\lambda}=2T_{\lambda}$. We can now replace the expressions for the contortion components (\ref{eq:ansatz}) in the expression (\ref{eq:Jcomp}) or derive this induced four-current correction directly from the (effective) previous Lagrangian in (\ref{eq:Lagcom}),
to arrive at the result
\begin{equation}
J^{k}=\lambda \left[4(f^{2}+h^{2})A^{k}(t)+f(t)\tilde{F}_{bc}\epsilon^{kbc}-2h(t)\tilde{F}^{0k}\right].
\end{equation}

Then, if $\vec{A}(t)\neq 0$, it obeys the following Maxwell-like equations in the cosmological (FLRW) context
\begin{equation}
\label{Maxeqsreview}
\ddot{A}_{k}+H\dot{A}_{k}=\lambda^{-1}(j_{k}+J_{k}) \ ,
\end{equation}
where $H=\dot{a}/a$ is the Hubble factor and  overdots denote time derivatives. Eq. (\ref{Maxeqsreview}) is an extended Proca-like equation with torsion contributing to an effective mass for the photon. We point out that for homogeneous fields we have $\tilde{F}_{bc}=0$ and  we can furthermore set, for simplicity, $j_{k}=0$ (on average) for comoving observers, in agreement with the cosmological principle.

\subsection{Fermionic (spinor) fields}

As for the Dirac sector, from Eq. (\ref{DiracLagv2}) and using the constraints on the contortion tensor (\ref{eq:Kansatz}), we arrive at
\begin{eqnarray}
\mathcal{L}_{\rm D}&=&\tilde{\mathcal{L}}_{\rm D}+
\dfrac{\hbar}{4}K_{abc}(\epsilon^{cab0}\bar{\psi}\gamma_{0}\gamma^{5}\psi+\epsilon^{cabd}\bar{\psi}\gamma_{d}\gamma^{5}\psi)
	\nonumber \\
&&+\dfrac{\hbar}{4}K_{0bc}(\epsilon^{c0bd}\bar{\psi}\gamma_{d}\gamma^{5}\psi-\epsilon^{cb0d}\bar{\psi}\gamma_{d}\gamma^{5}\psi) \ ,
\end{eqnarray}
which reads explicitly
\begin{equation}
\mathcal{L}_{\rm D}=\tilde{\mathcal{L}}_{\rm D}+
\dfrac{\hbar}{4}f(t)\epsilon_{abc}(\epsilon^{cab0}\bar{\psi}\gamma_{0}\gamma^{5}\psi+\epsilon^{cabd}\bar{\psi}\gamma_{d}\gamma^{5}\psi) \ .
\end{equation}
Alternatively, one could consider the
axial vector part of torsion (\ref{eq:axialpart}) in the ansatz (\ref{eq:ansatz}):
\begin{equation} \label{eq:axialant}
\breve{T}^{\lambda}=\dfrac{1}{6}f(t)\epsilon^{\lambda abc}\epsilon_{abc} \ ,
\end{equation}
and combine it with Eq. (\ref{DiracLag3}) to arrive at
\begin{equation}
\mathcal{L}_{\rm D}=\tilde{\mathcal{L}}_{\rm D}+\dfrac{\hbar}{4}f(t)\epsilon^{\lambda abc}\epsilon_{abc}\bar{\psi}\gamma_{\lambda}\gamma^{5}\psi \ .
\end{equation}
Since here $a,b,c=1,2,3$, we get
\begin{equation}
\mathcal{L}_{\rm D}=\tilde{\mathcal{L}}_{\rm D}+\dfrac{3\hbar}{2}f(t)\bar{\psi}\gamma_{0}\gamma^{5}\psi \ .
\end{equation}
The corresponding Dirac equation for this Lagrangian is
\begin{equation} \label{eq:Diracf}
i\hbar \gamma^{\mu}\tilde{D}_{\mu}\psi-m\psi =- \dfrac{3\hbar}{2}f(t)\gamma_{0}\gamma^{5}\psi \ .
\end{equation}

In the cosmological context, by performing a conformal transformation $g_{\mu\nu}=a^{2}(\eta)\eta_{\mu\nu}$, one can use the identity $\gamma^{\mu}\tilde{D}_{\mu}\psi=a^{-5/2}(\eta)\gamma^{\mu}\partial_{\mu}( a^{3/2}(\eta)\psi)$, to arrive at the Dirac equation in an FLRW cosmological background
\begin{equation} \label{eq:DiracFLRM}
i\hbar \gamma^{0}\chi'-a(\eta)\left(m-\dfrac{3\hbar}{2}f(\eta)\gamma_{0}\gamma^{5}\right)\chi=0 \ ,
\end{equation}
where $\chi(\eta)\equiv a^{\frac{3}{2}}(\eta)\psi$ and $\bar{\chi}(\eta)\equiv a^{\frac{3}{2}}(\eta)\bar{\psi}$, while now time derivatives are performed with respect to the conformal time $\eta$.
We can also write the above equation as the linear non-autonomous dynamical system
\begin{equation}
\vec{\chi}'=\bold{A}(\eta)\vec{\chi} \ ,
\end{equation}
where $\vec{\chi}$ is a four-spinor and $\bold{A}$ is the matrix with components $(\bold{A})_{CD}=-\frac{i}{\hbar}a(\eta)\left(m\gamma_{CD}^{0}-\dfrac{3\hbar}{2}f(\eta)\gamma^{5}_{CD}\right)$.
This equation can be solved, in principle, upon specification of the torsion function and the evolution of the scale factor.

\subsection{Gravitational field equations}

The equations above governing the dynamics of homogeneous fermionic and bosonic fields are coupled to the gravitational (Friedmann) equations whose explicit form we now derive. These equations can be conveniently written in the following form
\begin{equation} \label{eq:Eineq}
\tilde{G}_{\mu\nu}=\kappa^{2}\left(T^{\rm{pfluid}}_{\mu\nu}+U^{\rm tor}_{\mu\nu}+\Pi^{\rm M}_{\mu\nu}+\Sigma_{\mu\nu}^{D}\right),
\end{equation}
where $\tilde{G}_{\mu\nu}$ is the usual Einstein tensor of the Levi-Civita connection. Let us analyze each piece on the right-hand side of these equations separately. Note that the first term $T^{\rm{pfluid}}_{\mu\nu}=\text{diag}(-\rho,p,p,p)$ is just the standard energy-momentum tensor of a perfect fluid.

The second term is related to contortion, and defined as
\begin{equation} \label{eq:Contten}
U^{\rm tor}_{\mu\nu}=-\dfrac{2}{\sqrt{-g}}\dfrac{\delta (C\sqrt{-g})}{\delta g^{\mu\nu}} \ ,
\end{equation}
with the expression
\begin{equation} \label{eq:Cgen}
C=-\dfrac{1}{2\kappa^2}
\Big(K^{\alpha}{}_{\lambda\alpha}K^{\nu\lambda}{}_{\nu}+K^{\alpha\lambda\beta}K_{\lambda\beta\alpha}\Big) \ ,
\end{equation}
in agreement with the Einstein-Hilbert action in a RC spacetime, where the curvature scalars are related as
\begin{equation} \label{eq:Ricciscalar}
R=\tilde{R}-2\tilde{\nabla}^{\lambda}K^{\alpha}{}_{\lambda\alpha} +g^{\beta\nu}\left(K^{\alpha}{}_{\lambda\alpha}K^{\lambda}{}_{\beta\nu}-K^{\alpha}{}_{\lambda\nu}K^{\lambda}{}_{\beta\alpha}\right) .
\end{equation}
In the expression for $C$ we neglected the total derivative term $2\tilde{\nabla}^{\lambda}K^{\alpha}{}_{\lambda\alpha}$, since it does not contribute to the field equations.
After some algebra, this leads to the explicit expression
\begin{eqnarray}
U^{\rm tor}_{\mu\nu}&=&2\kappa^{-2}\Big[K^{\;\;0a}_{(\nu}K_{0a\vert\mu)}+K^{\;\;ab}_{(\nu}K_{ab\vert\mu)}+K^{a\;\;b}_{\;\;(\nu}K_{\mu)ba}
	\nonumber \\
&&+K^{a0}_{\;\;\;(\nu}K_{0\vert\mu)a}
	+K^{ab}_{\;\;\;(\nu}K_{b\vert\mu)a}\nonumber \\
&&+2K_{(\mu\;\;\;\nu)}^{\;\;\;\lambda}K_{\lambda}+K_{\mu}K_{\nu}\Big]+Cg_{\mu\nu} \ .
\end{eqnarray}
Using now the ansatz (\ref{eq:ansatz}) we have $K^{\lambda}K_{\lambda}=64h^{2}(t)$ and $K^{\alpha\lambda\beta}K_{\lambda\beta\alpha}=6f^{2}(t)-16h^{2}(t)$, which turns Eq. (\ref{eq:Cgen}) into
\begin{equation} \label{eq:Cexp}
C=-\dfrac{1}{2\kappa^2}
\Big(6f^{2}(t)+48h^{2}(t)\Big) \ .
\end{equation}
Associating energy density and pressure contents to the temporal and spatial components of the tensor (\ref{eq:Contten}), like in a standard perfect fluid, one finds an energy density
\begin{equation}
\rho^{\rm tor}=-\kappa^{-2}\left(3f^{2}(t)-160h^{2}(t)\right) \ ,
\end{equation}
while the pressure terms ($p\delta^{a}_{b}=-U^{a}_{b}$) are
\begin{equation}
(U^{\text{tor}})^{a}_{b}=-\kappa^{-2}\left(104h^{2}-9f^{2}(t)\right)\delta^{a}_{b} \ .
\end{equation}
As the simplest case, setting $h(t)=0$ we get the equation of state $w^{\rm tor} \equiv p^{\rm tor}/\rho^{\rm tor}=3$.

Regarding the two last contributions in Eq. (\ref{eq:Eineq}), they are defined as
\begin{eqnarray}
\Pi^{\rm M}_{\mu\nu}&=&-\dfrac{2}{\sqrt{-g}}\dfrac{\delta (\sqrt{-g}\mathcal{L}^{\rm M}_{\rm corr})}{\delta g^{\mu\nu}}\,,  \\
\Sigma^{\rm D}_{\mu\nu}&=&-\dfrac{2}{\sqrt{-g}}\dfrac{\delta (\sqrt{-g}\mathcal{L}^{\rm D}_{\rm corr})}{\delta g^{\mu\nu}} \,,
\end{eqnarray}
respectively.
Let us deal first with $\Pi^{\rm M}_{\mu\nu}$. It represents the effective energy-momentum contribution from the coupling between torsion and the electromagnetic (radiation) field. By keeping only terms quadratic in the torsion functions, we arrive at
\begin{eqnarray}
\Pi_{\mu\nu}^{\rm M}&=&-4\lambda\Big[(T^{\lambda\;\;\;\beta}_{\;\;(\mu}T_{\gamma\vert\nu)\beta}+T^{\lambda\alpha}_{\;\;\;\;(\mu}T_{\gamma\alpha\vert\nu)})A_{\lambda}A^{\gamma}\nonumber\\
&&+T^{\;\;\alpha\beta}_{(\mu}T_{\gamma\alpha\beta}A_{\nu)}A^{\gamma}
	\nonumber \\
&&+T^{\lambda\alpha\beta}T_{(\mu\vert \alpha\beta}A_{\lambda}A_{\nu)}\Big]+\mathcal{L}^{\rm M}_{\rm corr}g_{\mu\nu} \ .
\end{eqnarray}
The energy density and pressure terms associated to this tensor read
\begin{equation}
\rho^{\rm M}=-\lambda\left[A^{a}A_{a}(6h^{2}(t)-2f^{2}(t))-2h(t)\tilde{F}_{0b}A^{b}\right],
\end{equation}
\begin{eqnarray}
&&(\Pi^M)_{ij}=-4\lambda\left[(2f^{2}(t)+6h^{2}(t))A_{i}A_{j}-4h(t)\tilde{F}_{0(i}A_{j)}\right]
	\nonumber \\
&&\qquad +\lambda\left[(h^{2}(t)-6f^{2}(t))A^{k}A_{k}-2h(t)\tilde{F}_{0c}A^{c}\right]g_{ij}\,,
\label{pressba}
\end{eqnarray}
respectively, where we already considered homogeneous fields ($\tilde{F}_{bc}=0$). The second term in expression (\ref{pressba}) is compatible with isotropy, whereas the first term can introduce an anisotropic pressure/stress. If we set $h(t)=0$ we get 
\begin{equation}
\rho^{\rm M}=2\lambda f^{2}(t)\vec{A}^{2},
\end{equation}
\begin{equation}
(\Pi^M)_{m}^{k}=-8\lambda f^{2}(t)A^{k}A_{m}-6\lambda f^{2}(t)\vec{A}^{2}\delta^{k}_{m},
\label{pressba2}
\end{equation}
and if we neglect the anisotropic term in the equation above we obtain the isotropic pressure $p^{M}=6\lambda f^{2}(t)\vec{A}^{2}$ and the equation of state $w^{\rm M} \equiv p^{\rm M}/\rho^{\rm M}=3$. For the Friedmann equations we will be neglecting the energy density contribution from the minimal coupling of electromagnetism to fermions, assuming that on average $j^{\mu}\approx 0$, and take only the above energy density due to the coupling of the bosonic field to torsion.

As for the $\Sigma^{\rm D}_{\mu\nu}$ piece, which represents the contribution from Dirac fields, it can be computed as
\begin{equation}
\Sigma^{\rm D}_{\mu\nu}=3\breve{T}^{\lambda}\breve{s}_{\lambda}g_{\mu\nu}-12\breve{T}_{(\mu}\breve{s}_{\nu)} \ ,
\end{equation}
and since from Eq. (\ref{eq:axialant}) one has that $\breve{T}^{0}=f(t)$ ($\breve{T}^{m}=0$), then $\breve{T}^{\lambda}$ is necessarily time-like, as a consequence of the cosmological principle, and we obtain the associated densities and pressures as
\begin{eqnarray}
\rho^{\rm D}&=&-12\breve{T}_{0}\breve{s}_{0}+3\breve{T}^{0}\breve{s}_{0}=-9f(t)\breve{s}_{0},
	\\
(\Sigma^{\rm D})^a_{b}&=&3\breve{T}^{0}\breve{s}_{0}\delta^{a}_{b}=3f(t)\breve{s}_{0}\delta^{a}_{b}=-p^{\rm D}\delta^{a}_{b},
\end{eqnarray}
for the energy and pressure terms, respectively. In this case we have $w^{\rm D} \equiv p^{\rm D}/\rho^{\rm D}=1/3$.

A few remarks on these derivations are in order. First, we are not considering any condition on $\breve{s}^{\lambda}$ to be space-like (and therefore $u^{\alpha}\breve{s}_{\alpha}=0$ for comoving observers), but we are rather focusing on the cosmological principle restriction for the torsion components. Second, we have not used Cartan's equations yet. Should we use them, then in the simplest case where torsion is due to the spin tensor of fermions one would have $\breve{T}^{\lambda}=-\kappa^{2}\breve{s}^{\lambda}/2$ and therefore, if the cosmological principle is also taken into consideration, one is led to conclude that $\breve{s}^{\lambda}$ has to be time-like as well. Finally, we emphasize that in this subsection we are considering fermionic and bosonic fields propagating in the cosmological torsion background. Under a suitable averaging procedure these fields can then contribute as perfect fluids to the cosmological Friedmann equations but torsion can be seen here as an extra (external) homogeneous and isotropic tensor field that enriches the background spacetime geometry. We are neglecting the source of torsion, for simplicity.

\subsection{Friedmann equations}

We have all the elements ready to discuss the Friedmann equations for the ECDM model according to the implementation of the cosmological principle. The first such equation reads
\begin{eqnarray}
\label{Friedmannreview}
\dfrac{\dot{a}^{2}}{a^{2}}&=&\frac{\kappa^2}{3}(\rho+\rho^{\rm corr})-\frac{k}{a^{2}} \ ,
\end{eqnarray}
where the correction to the usual energy density of the perfect fluid $\rho$ is spelled out as $\rho^{\rm corr}=\rho^{\rm tor}+\rho^{\rm M}+\rho^{\rm D}$. It can be cast, as a function of the background (external) torsion and the fundamental matter fields ($\vec{A},\psi,\bar{\psi}$) as
\begin{eqnarray} \label{eq:rhocorr}
\rho^{\rm corr}&=&-f^{2}\left(\dfrac{3}{\kappa^{2}}-2\lambda\vec{A}^{2}\right)
-h^{2}\left(6\lambda \vec{A}^{2}-\dfrac{160}{\kappa^{2}}\right)\nonumber \\
&&-9f(t)\breve{s}_{0}+2h\tilde{F}_{0b}A^{b} \ ,
\end{eqnarray}
where  $\vec{A}^{2} \equiv A^{c}A_{c}$. The Friedmann equation above, together with the Maxwell equation in (\ref{Maxeqsreview}) and the Dirac equation (\ref{eq:DiracFLRM}) form the resulting dynamical system.

The generalized continuity equation, $\tilde{\nabla}^{\mu} T_{\mu\nu}^{\text{total}}=0$, in the FLRW background, reads
\begin{equation} \label{eq:conseq}
\dot{\rho}+3H\rho(1+w^{\rm rad})=-\left[ \dot{\rho}^{\rm corr}+3H\left( \rho^{\rm corr}+p^{\rm corr}\right)\right],
\end{equation}
where $w^{\rm rad}=1/3$ and $\rho$ refer to the usual relativistic matter (radiation) term in the early Universe, and one can set the right-hand side equal to zero (independent conserved fluid components), with the total pressure correction term (neglecting any anisotropic contributions) given by
\begin{eqnarray}
 p^{\rm corr}\delta^{k}_{m}&=&-f^{2}\left(\dfrac{9}{\kappa^{2}}-6\lambda\vec{A}^{2}\right)\delta^{k}_{m}
	-3f(t)\breve{s}_{0}\delta^{k}_{m}
	\nonumber \\
&&  + h^{2}\left(\dfrac{104}{\kappa^{2}} - \lambda\vec{A}^{2}\right)\delta^{k}_{m} + 2h\tilde{F}_{0b}A^{b}\delta^{k}_{m} \,.
\end{eqnarray}

To solve the full dynamics some simplifications must be made. For instance, one can take the case in which the torsion functions are constrained as $h(t)=0$ and $f(t)\neq 0$, which, as we will show later (see Sec. \ref{sec:Cartan}), follows naturally from the Cartan equations. Then, in that case where we take $\rho^{\rm tor}$, $\rho^{\rm D}$ and $\rho^{\rm M}$ as independent fluid components, representing the energy density of torsion self-interactions, torsion-fermions and torsion-bosons interactions, respectively, we  have: $w^{\rm tor}=3$, $w^{\rm D}=1/3$, $w^{\rm M}=3$, respectively.
Now, should we consider the hypothesis that these fluid components are independently conserved, then as expected for conserved barotropic fluids we would have $\rho^{\rm tor}\sim  a^{-12}$ (with $\rho^{\rm tor}<0$), $\rho^{\rm D}\sim  a^{-4}$ and $\rho^{\rm M}\sim a^{-12}$ (with $\rho^{\rm M}<0$), together with the usual radiation term $\rho\sim a^{-4}$.

In these conditions, the solution to the dynamics is inferred from the Friedmann equation (\ref{Friedmannreview}), which allows to compute the family of possible trajectories in the plane $(a,\dot{a})$. If we take also $\rho^{\rm D}<0$ (which will turn out to be valid, as we will see after using the Cartan equations), these trajectories show that the dynamics is that of a non-singular Universe, with very strong torsion effects avoiding the singularity, and replacing it by a minimal value of the scale factor, from which the Universe undergoes a period of early accelerated expansion followed by a period of decelerated expansion.
Nevertheless, the correct treatment should come from considering the torsion tensor as function of the matter fields, using the Cartan equations. This is what we will do in the last part of this paper.

Before going that way it should be pointed out that, regardless of what the source of torsion is, if the gravity theory introduces an effective (torsion) fluid correction to the Friedmann equations, and assuming that this can be decomposed as $\rho^{\rm corr}\sim \rho_{s}+\rho_{s-A}$, where $\rho_{s}$  is due to a spin-spin interaction of fermions and $\rho_{s-A}$ is the interaction between fermions and bosons (for instance $\rho_{s-A}\sim\rho^2_{s}A^{2}$), both effects induced by torsion, then assuming that $\rho_{s-A}\approx \Sigma $  is constant, the Friedmann equation (\ref{Friedmannreview}) becomes quite simple:
\begin{equation}
\left(\dfrac{\dot{a}}{a}\right)^2=\frac{\kappa^2}{3}(\rho+\rho^{s}(a)+\Sigma)-\frac{k}{a^{2}} \ .
\end{equation}
Furthermore, assuming that $\rho^{s}\sim a^{-6}$ (which is a reasonable assumption if $\rho^{s}\sim n^{2}$, where $n$ is the number density of fermions), with $\rho^{s}<0$, we get
\begin{equation}
H^{2}(a)=H_{0}^{2}\left[ \Omega^{\rm rad}_{\rm 0}\left(\dfrac{a_{\rm 0}}{a}\right)^{4}+\Omega^{\rm s}_{\rm 0}\left(\dfrac{a_{\rm 0}}{a}\right)^{6}+\dfrac{\kappa^{2}\Sigma}{3H_{0}^{2}}\right] -\frac{k}{a^{2}H_{0}^{2}},
\end{equation}
with $\Omega^{\rm s}_{\rm 0}<0$ and as usual the subscript $0$ refers to the present value for the corresponding cosmological parameter, and $\Omega^{j}_{\rm 0}=8\pi G\rho^{j}_{0}/3H_{0}$, for the fluid species $j$.
Then, as shown in \cite{Cabral:2020mst}, the resulting dynamics yields a bounce in the early Universe for the three cases $k=-1,0,1$. Moreover, if $\Sigma<0$, a future cosmological bounce is also found, as given by the zeroes of  $H(a)=\pm\sqrt{{\rho}^{\rm eff}(a)-k/a^{2}}$, and since $H(a)$ cannot become imaginary, the solution can transit from the positive branch to the negative one, corresponding to a transition from decelerated expansion into accelerated contraction. In the positive branch, the solutions show an early accelerated expansion phase from a bounce at the minimal value of the scale factor, until $H(a)$ reaches a local maximum, followed by the decelerated expansion phase until the future bounce. In the negative branch, i.e, the contracting phase, after a period of accelerated contraction the Hubble parameter reaches a local minimum and starts increasing in a decelerated contraction phase due to the spin-spin repulsive effects preventing the singularity to occur, and a new bounce and subsequent accelerated expansion establish a cyclic cosmology. 

On the other hand, if $\Sigma>0$ the dynamical system describes a non-singular cosmology again with a period of early accelerated expansion followed by a period of decelerated expansion, until the Hubble parameter stabilizes (asymptotically) at a fixed constant value. This is valid for all geometries, and the inclusion of the term $\rho_{s-A}\approx \Sigma $ leads therefore to an effective (positive) cosmological constant and the corresponding period of late-time accelerated expansion. The addition to this scenario of the (dust) matter term representing baryonic and dark matter fluids does not change this general dynamics. This example illustrates well how a simple model as the ECDM with minimal couplings between torsion and the matter fields can give rise to such remarkable solutions without the need for any inflaton, quantum gravity or dark energy, yielding in fact a unique classical solution that is non-singular, includes an abrupt early acceleration period, and a late-time acceleration period.

\subsection{Cartan relations and the cosmological principle} \label{sec:Cartan}

Writing the total matter Lagrangian as $\mathcal{L}=C+\mathcal{L}^{\rm M}_{\rm corr}+\mathcal{L}^{\rm D}_{\rm corr}$, where $C$ is given by Eq. (\ref{eq:Cexp}) and the other two terms by the expressions
\begin{eqnarray} \label{eq:Lagcomb}
\mathcal{L}^{\rm M}_{\rm corr}&=&\lambda\Big[2\left(f^{2}(t)+h^{2}(t)\right)A^{a}A_{a}\nonumber \\
&&+f(t)\epsilon^{abc}\tilde{F}_{bc}A_{a}-2h(t)\tilde{F}_{0b}A^{b}\Big] \,,
\end{eqnarray}
and
\begin{equation}
\mathcal{L}_{\rm corr}^{\rm D}=\dfrac{3\hbar}{2}f(t)\bar{\psi}\gamma_{0}\gamma^{5}\psi \,,
\end{equation}
and varying it with respect to $f(t)$ and $h(t)$ we find the Cartan equations in this case as
\begin{equation} \label{eq:Cart1}
2f(t)\left(2\lambda\vec{A}^{2}-\dfrac{3}{\kappa^2}\right)+\dfrac{3\hbar}{2}\bar{\psi}\gamma_{0}\gamma^{5}\psi+\lambda\tilde{F}_{bc}\epsilon^{abc}A_{a}=0,
\end{equation}
and
\begin{equation}
h(t)\left(4\lambda\vec{A}^{2}-2\lambda\tilde{F}_{0b}A^{b}-\dfrac{48}{\kappa^2}\right)=0 \ ,
\end{equation}
respectively, and we can set $h(t)=0$. Since for homogeneous fields one has that  $\tilde{F}_{ab}=0$ then Eq. (\ref{eq:Cart1}) allows us to solve for $f(t)$ as
\begin{equation} \label{eq:f(t)gen}
f(t)=-\dfrac{3\kappa^2\hbar}{2}\bar{\psi}\gamma_{0}\gamma^{5}\psi\left(4\lambda\kappa^2\vec{A}^{2}-6\right)^{-1} \ .
\end{equation}
If we now substitute this torsion function back into the cosmological equations for the Maxwell (\ref{Maxeqsreview}) and Dirac fields (\ref{eq:DiracFLRM}) and in the Friedmann equation (\ref{Friedmannreview}),  we arrive at the final system of equations for the matter fields ($A_{\mu},\psi,\bar{\psi}$) and the scale factor $a$. By doing this one obtains non-linear equations for both Maxwell and Dirac fields with explicit non-minimal couplings between fermions and bosons.

Some specific scenarios can be now considered:

i)
For instance, in the ansatz of $A_{\mu}=(\phi,0,0,0)$, Eq.(\ref{eq:f(t)gen}) becomes
\begin{equation}
f(t)=\dfrac{\kappa^{2}\hbar}{4}\bar{\psi}\gamma_{0}\gamma^{5}\psi \ ,
\end{equation}
therefore in terms of the conformal time this torsion function reads
\begin{equation}
f(\eta)=\dfrac{\kappa^{2}\hbar}{4}(\bar{\chi}\gamma_{0}\gamma^{5}\chi)a^{-3}(\eta) \ ,
\end{equation}
which yields the (non-linear) Dirac equation (\ref{eq:DiracFLRM})
\begin{equation}
i\hbar \gamma^{0}\chi'-a(\eta)m\chi+\dfrac{3\kappa^{2}\hbar^{2}}{8}a^{-2}(\eta)\left(\bar{\chi}\gamma_{0}\gamma^{5}\chi\right)\gamma_{0}\gamma^{5}\chi =0 \ ,
\end{equation}
and the correction (\ref{eq:rhocorr}) to the  Friedmann equations becomes
\begin{eqnarray}
\rho^{\rm corr}&=&-\dfrac{3}{\kappa^{2}}f^{2}(t)-9f(t)\breve{s}_{0}\nonumber \\
&=&-\dfrac{21\kappa^{2}\hbar^{2}}{16}(\bar{\psi}\gamma_{0}\gamma^{5}\psi)\bar{\psi}\gamma_{0}\gamma^{5}\psi <0 \ .
\end{eqnarray}
In this case the bosonic field does not contribute with extra terms to the dynamics, besides the usual radiation term. Moreover, assuming that  $(\bar{\psi}\gamma_{0}\gamma^{5}\psi)^{2}\sim n^{2}\sim a^{-6}$, with $n$ being the fermionic number density\footnote{More rigorously, one should consider the corresponding expectation value $\left<(\bar{\psi}\gamma_{0}\gamma^{5}\psi)^{2}\right>\sim a^{-6}$.}, this correction can be written as
\begin{equation}
\rho^{\rm corr}=-\dfrac{21\kappa^{2}\hbar^{2}}{16}\beta_{s}a^{-6},
\end{equation}
where $\beta_{s}$ is a constant.

The cosmological solutions in this case are easy to obtain, and correspond to a non-singular bouncing behaviour due to the strong (repulsive) spin-spin effects induced by torsion in the very early-Universe. This non-singular behaviour is present in all possible spatial geometries ($k=-1,0,1$) including a period of early acceleration followed by the usual deceleration expansion phase of Friedmann models (without dark energy/cosmological constant).

ii)
On the other hand, if we set $\vec{A}\neq 0$, then  Eq.(\ref{eq:f(t)gen}) replaced in (\ref{eq:rhocorr}) yields the correction to the Friedmann equations
\begin{equation}
\rho^{\rm corr}=-\dfrac{63\kappa^{2}\hbar^{2}}{16}(\bar{\psi}\gamma_{0}\gamma^{5}\psi)^{2}\left(3-2\lambda\kappa^{2}\vec{A}^{2}\right)^{-1}<0 \ ,
\end{equation}
and using the same assumption as in the previous example it becomes
\begin{equation}
\rho^{\rm corr}=-\dfrac{63\kappa^{2}\hbar^{2}}{16} \left(3-2\lambda\kappa^{2}\vec{A}^{2}\right)^{-1}\beta_{s}a^{-6} \ ,
\end{equation}
where we recall that $\vec{A}^{2}\equiv A^{j}A_{j}<0$. The corresponding  Dirac-Hehl-Data equation (\ref{eq:DiracFLRM})  becomes
\begin{eqnarray}
i\hbar \gamma^{0}\chi'-a(\eta)m\chi &=&
-\dfrac{3\kappa^{2}\hbar^{2}}{8\left(4\lambda\kappa^{2}\vec{A}^{2} -6\right)} \times
	\nonumber \\
&&\times a^{-2}(\eta)(\bar{\chi}\gamma_{0}\gamma^{5}\chi)\gamma_{0}\gamma^{5}\chi \ .
\end{eqnarray}

Finding solutions in this case is less trivial as it involves a non-minimal coupling between the fermionic spin-spin energy interaction and the bosonic field. Due to this, it is not straightforward to derive the evolution of the bosonic potential from the Maxwell equations (\ref{Maxeqsreview}), using Eq. (\ref{eq:f(t)gen}), which are non-linear and coupled to Friedmann equations.
From the general expression for the pressure functions obtained above and neglecting anisotropic stresses one can prove that  this ``fluid" component obeys
\begin{equation}
p^{\rm corr}\approx -\dfrac{45\kappa^{2}\hbar^{2}}{16}\dfrac{(\bar{\psi}\gamma_{0}\gamma^{5}\psi)^{2}}{3-2\lambda\kappa^{2}\vec{A}^{2}}\ ,
\end{equation}
so therefore the equation of state of this fluid is given by $w^{\rm corr}\equiv p^{\rm corr}/\rho^{\rm corr}=45/63 $.

In principle, this can be used in the corresponding (fluid component) conservation (\ref{eq:conseq}), given in this case by
\begin{equation}
\dfrac{d\rho^{\rm corr}}{da}+\dfrac{3}{a}\left(w^{\rm corr}(A)+1\right)\rho^{\rm corr}=0 \ ,
\end{equation}
to obtain $\rho^{\rm corr}\sim a^{-3(w^{\rm corr}+1)}\approx a^{-5,14}$.  Nevertheless, one should take into account that we have neglected anisotropic (off-diagonal) components of the energy-momentum tensors describing the torsion corrections. One can also take a more general approach, considering that $\rho^{\rm corr}<0$ and assuming that its relevant effect on the dynamics can be qualitatively modeled by a power-law
\begin{equation}
\rho^{\rm corr}\sim\xi+\varsigma a^{-n}+\mu a^{-m} \ ,
\end{equation}
with the exponents $m,n$ being real numbers and the constants $\xi,\varsigma<0$ (to fulfill the condition $\rho^{\rm corr}<0$ at all times), then the resulting dynamics is quite rich. Indeed, as long as one of the exponents is larger than the corresponding exponent of the usual radiation fluid, i.e, $m>4$ or $n>4$, we have a non-singular early cosmology. Moreover, if the other term survives in the late-time dynamics, decaying slower than $a^{-2}$, then it will give rise to a future bounce, where $H(a)=0$, and a similar cyclic behavior as that found in Ref. \cite{Cabral:2020mst} is expected.

Let us finally point out that, for both of the ansazte for the bosonic four-vector sector above, the resulting non-linear Dirac equations are of the Hehl-Data-Heisenberg type with cubic terms that change sign upon a charge conjugation transformation. Accordingly, the cosmological dynamics for fermions and anti-fermions is different, with interesting consequences to the matter/anti-matter asymmetries.

\subsection{Ricci scalar and the cosmological principle}

Taking into account the general expression (\ref{eq:Ricciscalar}) for the Ricci scalar of a RC spacetime geometry and using the expression
\begin{eqnarray}
\tilde{\nabla}_{\lambda} K^{\lambda}&=&\partial_{0}K^{0}+\dfrac{1}{\sqrt{-g}}\partial_{0}(\sqrt{-g})K^{0}
	\nonumber \\
&=&\dot{K^{0}}+3H(t)K^{0} \ ,
\end{eqnarray}
where in the last equality we have implemented a FLRW geometry, we arrive at
\begin{equation}
R=\tilde{R}+8\left[\dot{h}(t)+3H(t)h(t)\right]-6f^{2}(t)-48h^{2}(t) \ ,
\end{equation}
which for $h(t)=0$ reads
\begin{equation}
R=\tilde{R}-6f^{2}(t) \ .
\end{equation}
This way, if one considers that torsion scales as $T\sim a^{-3}$ (which is a reasonable assumption in the case of fermionic torsion, i.e, when $T\sim \breve{s}$), then we get
\begin{equation}
R=\tilde{R}-\dfrac{\alpha}{a^{6}} \ ,
\end{equation}
where $\alpha$ is a constant. Therefore, in the asymptotic limit $a \rightarrow \infty$ we find that both scalars coincide, while for a certain $a_{b}$ corresponding to the minimum value of the scale factor at the (early) cosmological bounce, a non-singular description of curvature is obtained, where
\begin{equation}
R\rightarrow \tilde{R}(a_{b})-\dfrac{\alpha}{a^{6}_{b}}, \qquad (a\rightarrow a_{b}) \ ,
\end{equation}
is finite. Let us point out that only for $\rho^{\rm corr}<0$ it is possible to get an early bounce (where $H(a_{b})=0$), that is to say, to prevent a singular behavior. It should be possible to prove in this type of models the geodesic completeness of the full set of solutions, which in principle suggests the interpretation of a previous contracting Universe phase.

\section{Conclusion and discussion} \label{secIV}

In this work we have addressed the implementation of the cosmological principle in a theory with torsion, namely, Einstein-Cartan gravity including minimal couplings with Dirac and Maxwell fields to the Riemann-Cartan geometry (ECDM model). In this theory, one finds generalized Dirac-like and Maxwell-like equations coupled to the background torsion, and by using the Cartan equations relating the torsion  tensor to the matter fields, one explicitly obtains both self-interactions (fermion-fermion and boson-boson) and non-minimal fermion-boson interactions.
The correct application of the cosmological principle to theories with torsion and, in particular, to the ECDM model, requires a careful and critical analysis.

The cosmological principle is motivated by the fact that in observational cosmology we observe a high degree of spatial isotropy both in the maps of the CMB radiation and in the observations of the distributions of clusters of galaxies at large scales. Then, assuming the laws of physics to be the same for any observer and that we do not constitute a preferred class of observers, we extrapolate these observations to the idea of the cosmological principle (spatial homogeneity and isotropy), valid for all comoving observers. In GR this gives rise to FLRW metrics representing the metric structure of the possible spatially maximally symmetric spacetimes and the perfect fluid form for the energy-momentum tensor.

On the other hand, since matter is known to be more accurately described by fundamental field theories, then typically it is quite reasonable to assume (spatially) homogeneous and isotropic vector, tensor or spinor fields in a cosmological context upon some appropriate average procedure. However, the imposition of the maximally symmetric spaces directly on the geometrical side is a much stronger requirement since it is imposing global spatial symmetries into the metric tensor that characterizes the local geometry (metric structure). Nevertheless, and although it is also reasonable to take the FLRW metric as a zeroth order approximation and consider the metric fluctuations only at the level of perturbation theory, it can be argued that the imposition of the cosmological principle to the torsion tensor right from the start has some advantages and some limitations. This follows from a careful analysis of the paradigm changes that are required to consistently interpret gauge theories of gravity with non-Riemann geometries.

In particular, the Einstein-Cartan theory is a simple example of a Poincare gauge theory of gravity with a RC spacetime geometry and, as such, several consequences are unavoidable, given the nature of the physical fields representing matter due to the richer RC geometry, and also as far as the connection between these matter fields and geometrical structures (torsion) is concerned. In particular, the approximation of a point-like particle is no longer valid, because its description from a multipolar expansion approach of matter fluids is not compatible with the generalized Bianchi identities in RC geometry \cite{Poplawski:2009su}. Therefore, non-Riemann geometries seem to suggest strong limitations in the classical picture of matter and are suitable to provide a more consistent description of microscopic gravity \cite{Obukhov} at classical and semi-classical regimes. Indeed, instead of considering geodesic equations  (valid for point-like particles) to study the effects of non-Riemannian geometries  \cite{Cembranos:2016xqx}, one should consider instead the fermionic field equations (generalized Dirac) and the bosonic field equations (generalized Maxwell or Proca-like), propagating in such geometries, or the equations of motion of extended test objects having the appropriate (Noether) current properties \cite{Obukhov:2014mka}. This supports the approach of taking Cartan's equations as valid also in the microscopic realm.

In view of the discussion above, for cosmological applications it seems more reasonable to impose the condition that the torsion tensor should be given as a function of the fundamental matter fields (via the spin tensor in Cartan equations) and take these fields to depend on time (to respect spatial homogeneity). Then a macroscopic averaging procedure should guarantee the isotropy requirement, as explicitly implemented in \cite{Cabral:2019gzh} by considering random fermionic spin distributions and random bosonic three-vector fields. Due to these considerations, we favour the cosmological models with Cartan's equations as microscopically meaningful, for example using fundamental fermionic spinor fields instead of a Weyssenhof fluid. Moreover, we should take the imposition of the cosmological principle to the torsion tensor as a zeroth-order approximation, having its limitations, that can be improved via perturbation theory or, more consistently, in fully non-homogeneous and anisotropic models.

In any case, in the present work we obtained the restrictions from the implementation of the cosmological principle on the torsion tensor components. We considered fermionic and bosonic fields propagating in the cosmological torsion background. These fields, under a suitable average procedure, contribute to the usual perfect fluid components in the cosmological Friedmann equations and also ``feel'' the torsion which can be considered as an extra (external) homogeneous and isotropic tensor field that enriches the background spacetime geometry. We derived the generalized Maxwell-like, Dirac-like and the Friedmann equations in this scenario. While the matter field equations are linear at this level of the analysis, when substituting the torsion function $f(t)$ as a function of the matter fields, using Cartan's equations (the latter derived from the effective total Lagrangian via variation with respect to the torsion functions), one arrives at a non-linear dynamics with non-minimal couplings in the matter sectors. From these, rich non-singular cosmological scenarios emerge which deserve more investigation, particularly with regards to small perturbations and the corresponding imprints in the CMB or in gravitational wave cosmological backgrounds from phase transitions in the early Universe.
In both the ECSK and the ECDM models the dominating torsion effects in the early Universe prevent an initial singularity and a simple interpretation of the solutions is that of a bounce from a previous contracting phase. At the bounce the curvature scalar is finite and one can respect geodesic completeness.

To conclude, a rigorous implementation of the cosmological principle in theories with torsion introduces some subtle modifications in their  cosmological dynamics, which has a relevant impact for their predictions, such as the appearance of non-minimal couplings, non-singular cosmologies, baryon-antibaryon asymmetries, etc. As limitations of our approach, let us point out that the energy-momentum tensor terms that were derived from the $U(1)$-breaking term in the Maxwell Lagrangian gives rise to an effective fluid description which can introduce anisotropic stresses. This should affect the dynamics via the Raychaudhuri equation and/or the conservation equation, though by simplicity in this work we neglected such effects. Similarly, by assuming the viewpoint that Cartan's equations should be approximately valid at a microscopic level our model would call for a more self-consistent cosmological approach, for instance within Bianchi spacetimes \cite{Capozziello:2017adv}. More work is thus necessary upon these theories to investigate their cosmological viability.

\section*{Acknowledgments}

FC is funded by the Funda\c{c}\~ao para a Ci\^encia e a Tecnologia (FCT, Portugal) predoctoral grant No.PD/BD/128017/2016.
FSNL acknowledges support from the FCT Scientific Employment Stimulus contract with reference
CEECIND/04057/2017.
DRG is funded by the \emph{Atracci\'on de Talento Investigador} programme of the Comunidad de Madrid, No.2018-T1/TIC-10431, and acknowledges further support from the projects FIS2014-57387-C3-1-P and FIS2017-84440-C2-1-P (MINECO/FEDER, EU), the project SEJI/2017/042 (Generalitat Valenciana), and PRONEX (FAPESQ-PB/CNPQ, Brazil). The authors also acknowledge funding from FCT projects No.UID/FIS/04434/2019 and No.PTDC/FIS-OUT/29048/2017.
FC thanks the hospitality of the Department of Theoretical Physics and IPARCOS of the Complutense University of Madrid, where part of this work was carried out. This article is based upon work from COST Actions CA15117  and CA18108, supported by COST (European Cooperation in Science and Technology).



\end{document}